\documentclass[a4paper]{PoS}

\title{Testing and Performance of UFFO Burst Alert \& Trigger Telescope}

\ShortTitle{Testing and Performance of UFFO Burst Alert \& Trigger Telescope}

\author{\speaker{Jakub \v{R}\'{\i}pa}%
         \thanks{On behalf of UFFO collaboration. This study was supported by Creative Research Initiatives (RCMST) of MEST/NRF.}\\
        Institute of Basic Science, Sungkyunkwan University\\
        2066 Seobu-ro, Suwon, 440-746, Korea\\
        E-mail: \email{ripa.jakub@gmail.com}}

\author{Min Bin Kim, Jik Lee, Il Huang Park, Ji Eun Kim, Heuijin Lim\\
Department of Physics, Sungkyunkwan University\\
2066 Seobu-ro, Suwon 440-746, Korea
}

\author{Soomin Jeong, Alberto Javier Castro-Tirado\\
Instituto de Astrof\'{\i}sica de Andaluc\'{\i}a (IAA-CSIC)\\
Glorieta de la Astronom\'{\i}a s/n, 18008, Granada
}

\author{Paul Henry Connell, Chris Eyles, Victor Reglero, Juana Maria Rodrigo\\
Image Processing Laboratory, University of Valencia\\
C/Catedr\'{a}tico Jos\'{e} Beltran, 2, 46980 Paterna (Valencia), Spain
}

\author{V.\ Bogomolov, M.\ I.\ Panasyuk, V.\ Petrov, S.\ Svertilov, I.\ Yashin\\
Skobeltsyn Institute of Nuclear Physics of Lomonosov Moscow State University\\
Leninskie Gory 119234, Moscow, Russia
}

\author{S.\ Brandt, C.\ Budtz-J\o rgensen\\
National Space Institute, Astrophysics, Technical University of Denmark\\
DK-2800 Kongens, Lyngby, Denmark
}

\author{Y.-Y.\ Chang$^{1}$, P.\ Chen$^{1}$, M.\ A.\ Huang$^{1,2}$, T.-C.\ Liu$^{1}$, J.\ W.\ Nam$^{1}$, M.-Z.\ Wang$^{1}$\\
$^{1}$Leung Center for Cosmology and Particle Astrophysics, National Taiwan University\\
1 Roosevelt Rd., Taipei 10617, Taiwan, R.O.C.
\\
\\
$^{2}$Department of Energy Engineering, National United University\\
2, Lienda Rd., Miaoli City, Taiwan, 36063, R.O.C.
}

\author{C.\ R.\ Chen\\
National Space Project Organization\\
8F., No.\ 9, Prosperity 1$^{\textrm{st}}$ Rd., Hsinchu Science Park, Hsinchu 30078, Taiwan, R.O.C.
}

\author{H.\ S.\ Choi\\
Korea Institute of Industrial Technology\\
89 Yangdaegiro-gil, Seobuk-gu, Cheonan-si, Chungcheongnam-do, 331-822, Korea
\\
\\
}

\author{S.-W.\ Kim\\
Center for Galaxy Evolution Research \& Department of Astronomy, Yonsei University\\
134 Shinchon-dong, Seoul 120-749, Korea
}

\author{K.\ W.\ Min\\
Department of Physics, Korea Advanced Institute of Science and Technology\\
291 Daehak-ro, Daejeon 305-701, Korea
}

\abstract{The Ultra-Fast Flash Observatory pathfinder (UFFO-p) is a new space mission dedicated to detect Gamma-Ray Bursts (GRBs) and rapidly follow their afterglows in order to provide early optical/ultraviolet measurements. A GRB location is determined in a few seconds by the UFFO Burst Alert \& Trigger telescope (UBAT) employing the coded mask imaging technique and the detector combination of Yttrium Oxyorthosilicate (YSO) scintillating crystals and multi-anode photomultiplier tubes. The results of the laboratory tests of UBAT's functionality and performance are described in this article. The detector setting, the pixel-to-pixel response to X-rays of different energies, the imaging capability for $< 50$\,keV X-rays, the localization accuracy measurements, and the combined test with the Block for X-ray and Gamma-Radiation Detection (BDRG) scintillator detector to check the efficiency of UBAT are all described. The UBAT instrument has been assembled and integrated with other equipment on UFFO-p and should be launched on board the Lomonosov satellite in late-2015.}

\FullConference{Swift: 10 Years of Discovery\\
		 2-5 December 2014\\
		 La Sapienza University, Rome, Italy}

\begin{document}

\section{Introduction and Description of UBAT}
\begin{figure}
\centering
    \makebox[0.45\linewidth][c]{\includegraphics[width=0.35\linewidth]{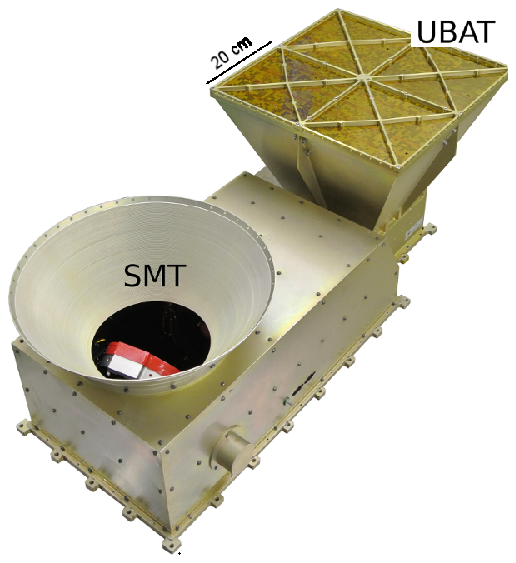}}   
    \makebox[0.45\linewidth][c]{\includegraphics[width=0.35\linewidth]{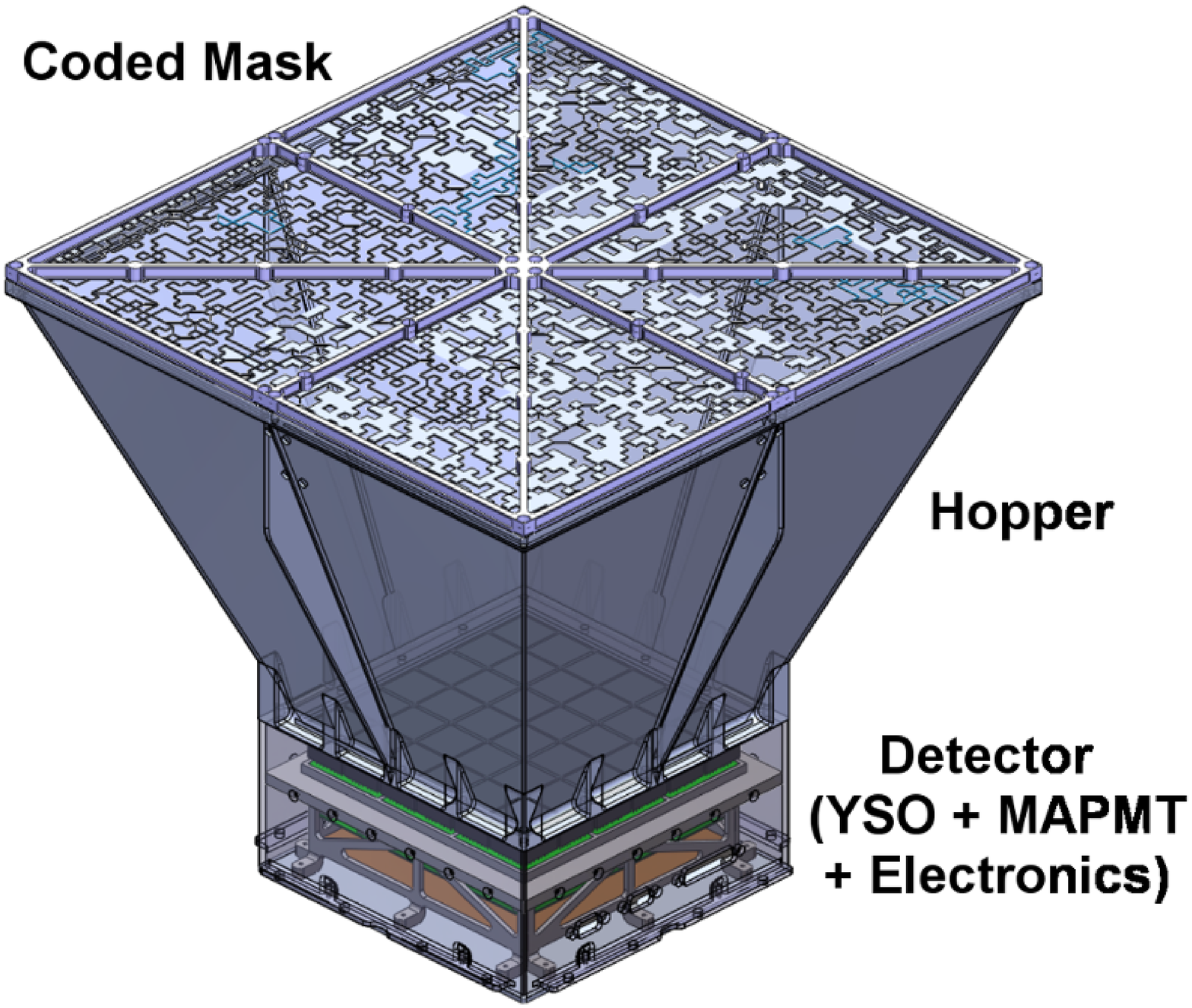}}
    \caption{
	    \textit{Left:} A photo shows the assembled UFFO-p with its two scientific instruments: UBAT and SMT.
	    \textit{Right:} A schematic view of UBAT is drawn. It consists of a coded mask, hopper, detector and readout electronics.
	    }
    \label{fig:uffo_ubat}
\end{figure}

The Ultra-Fast Flash Observatory pathfinder (UFFO-p) \cite{Chen11,Nam13,Park13} is a new space mission dedicated to detect Gamma-Ray Bursts (GRBs) and rapidly follow their afterglows to provide early optical/ultraviolet measurements. It consists of two scientific instruments. One is the UFFO Burst Alert \& Trigger telescope (UBAT) \cite{Jung11,Lee13,Chang14} which employs the coded mask technique providing X-ray/gamma-ray imaging observations in the energy range of $10-150$\,keV with half-coded field of view (FOV) $70.4^{\circ}\times70.4^{\circ}$ and angular resolution $\le10'$. The second instrument is the Slewing Mirror Telescope (SMT) \cite{Jeong13,Kim13} with the field of view of $17'\times17'$ and optical/ultraviolet range of $200-650$\,nm. It consists of a Ritchey-Chr\'{e}tien telescope with an Intensified Charge-Coupled Device in its focal plane together with a plane slewing mirror. In this article the laboratory tests of UBAT flight module (FM) (Figure~\ref{fig:uffo_ubat}) are briefly described. UBAT and SMT have been assembled and integrated with the control electronics on UFFO-p which is planned to be launched on the Lomonosov Moscow State University satellite \cite{Panasyuk11}.

UBAT is a wide-field X-ray imager which provides triggers of GRBs and determines their location via the coded mask technique. It consists of a coded mask, hopper, detector, and readout electronics. The coded mask is a 1\,mm thick plate of tungsten with a random pattern of $68\times68$ opened and closed tiles, having 44.5\,\% of open area. The size of each tile is 5.76\,mm$\times$5.76\,mm, i.e. double the detector pixels size. It is placed 280\,mm above the detector surface and it casts a mosaic shadow onto the detector. From the cross-correlation of the position of the projected shadow and the coded mask pattern one can determine the direction of the X-ray source \cite{Connell13}. The hopper supports the coded mask and provides shielding against the diffuse cosmic X-ray background. It is made of 2\,mm thick aluminium with 0.1\,mm thick layer of tungsten.
The detector comprises a Cesium-doped Yttrium Oxyorthosilicate (YSO) (Y$_{2}$SiO$_{5}$ : 0.2\,\% Ce) scintillator crystal array and Hamamatsu \textit{R11265-03-M64} Multi-Anode Photomultiplier Tubes (MAPMTs), together with the analogue and digital electronics. There are 36 MAPMTs and each of them has $8 \times 8 = 64$ channels, thus providing 2304 imaging pixels in total. On the top of each MAPMT there is a 8$\times$8 YSO crystal array. The volume of each single crystal is 2.68 (length) $\times$ 2.68 (width) $\times$ 3.0 (height)\,mm$^{3}$.

\section{Pixel-to-Pixel Response}
\begin{figure}
    \makebox[0.45\linewidth][c]{\includegraphics[width=0.4\linewidth]{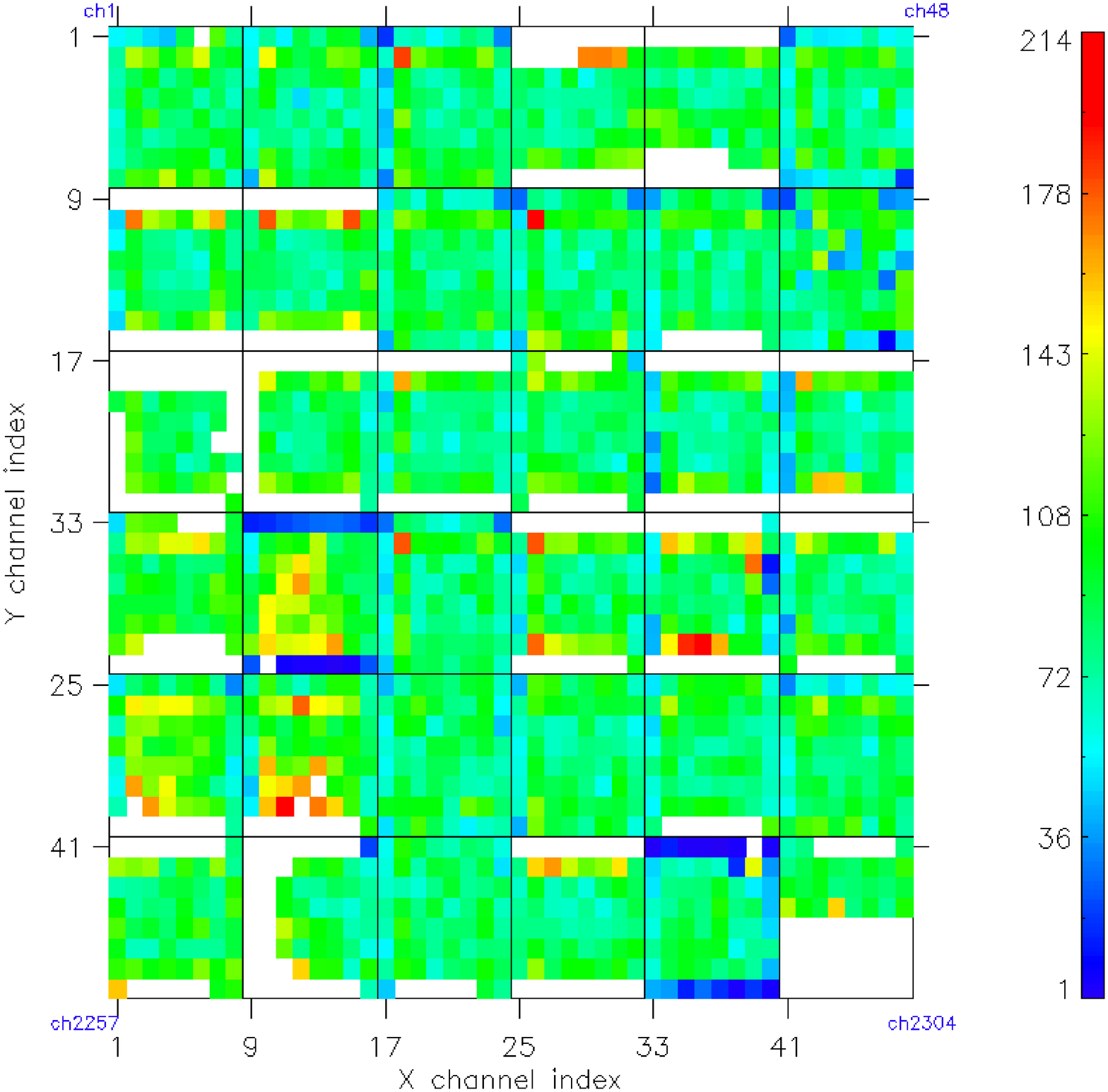}}   
    \makebox[0.53\linewidth][c]{\includegraphics[trim=-50pt -50pt 0pt 0pt, width=0.49\linewidth]{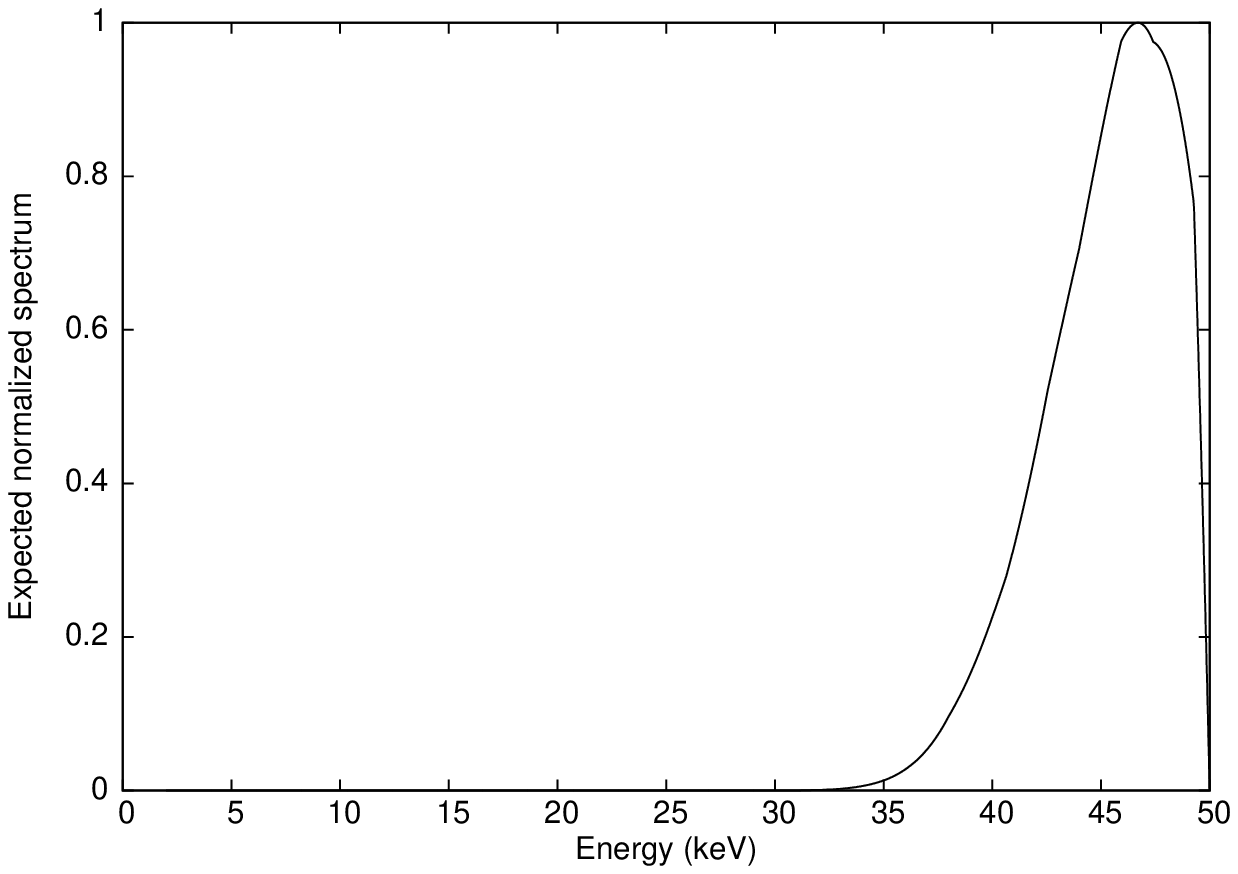}}
    \caption{
	    \textit{Left:} The UBAT FM detector response to X-rays of energies $\leq 50$\,keV is shown. The color scale indicates the total number of counts in each pixel. White color means zero counts at noisy detector pixels, which have been disabled noisy.
	    \textit{Right:} The expected normalized spectrum given by the radiation produced by the X-ray tube illuminating the detector after attenuation in the air and in the Cu filter is displayed.}
    \label{fig:hitmap-exp-spec-50kV}
\end{figure}

For the best imaging performance a uniform response of the detector is required because the imaging algorithm effectively subtracts any uniform background. In order to test the pixel-to-pixel response, to set the thresholds for each MAPMT and to determine the noisy pixels an X-ray tube was placed on axis in front of the detector without the coded mask attached. A tube with adjustable current and voltage of the accelerated electrons was used. It produced a broad-band Bremsstrahlung X-ray radiation up to 50\,keV. Figure~\ref{fig:hitmap-exp-spec-50kV} shows results for the following setup: source-detector distance of 8088\,mm, tube voltage (current) of 50\,kV (5.3\,$\mu$A), total exposure time of 186\,s and with a 1.6\,mm thick Cu filter placed in front of the tube to attenuate the flux.

The recorded data were processed off-line with analysis software which used a geometrical center hit-finder to eliminate the optical crosstalk caused by total reflections at the glass-vacuum (photo-cathode) interface \cite{Chang14}. There were 30\,460 registered counts due to the background and 137\,256 counts due to the source, and the area of the enabled detector pixels was 160.2\,cm$^2$ thus giving a count rate due to the source of 4.6 cnt\,s$^{-1}$\,cm$^{-2}$. Figure~\ref{fig:hitmap-exp-spec-50kV} shows the expected count spectrum given by the radiation produced by the X-ray tube after attenuation in the air and in the 1.6\,mm thick Cu filter. It was calculated from the measured X-ray spectrum given by the tube's manufacturer and considering the attenuation law.

\section{Test of Imaging}

\begin{figure}
     \makebox[0.4\linewidth][c]{\includegraphics[width=0.35\linewidth]{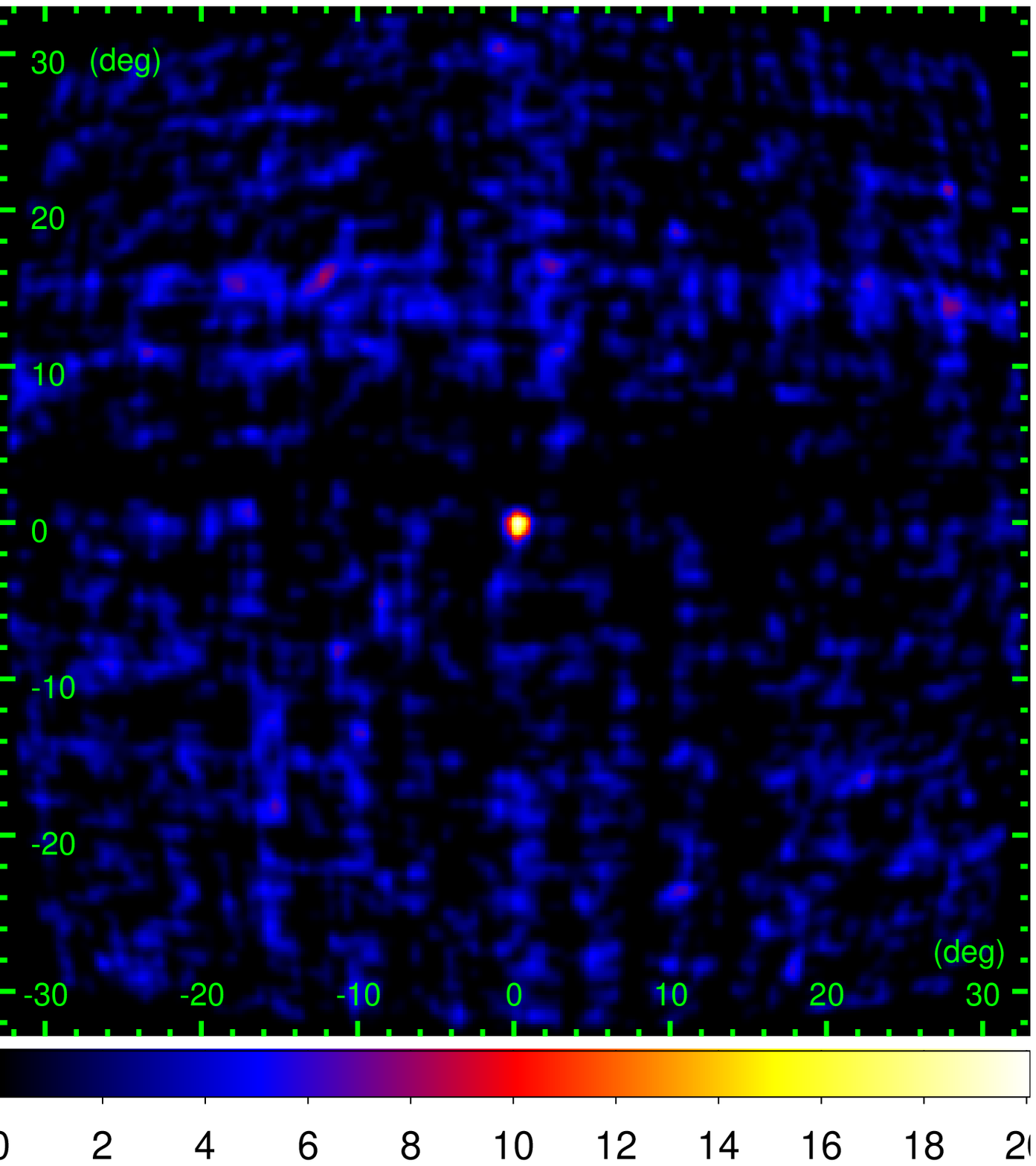}}   
    \makebox[0.6\linewidth][c]{\includegraphics[trim=-50pt -50pt 0pt 0pt, width=0.55\linewidth]{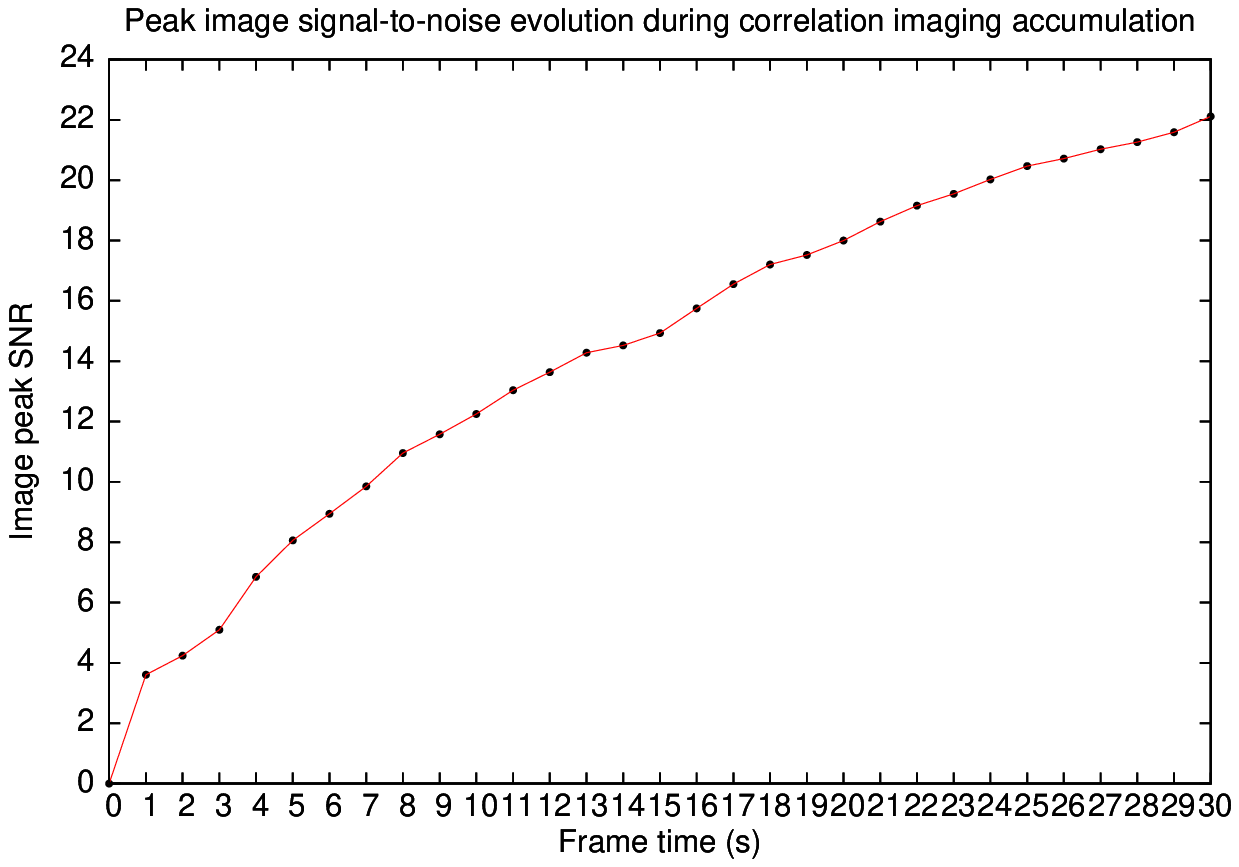}}
    \caption{
	    \textit{Left:} Reconstructed X-ray image as an SNR map. Color scale corresponds to the SNR value. The X-ray tube was placed on axis with voltage 50\,kV. The X-ray source is clearly visible in the middle of the FOV.
	    \textit{Right:} The image peak SNR evolution during the exposure time.}
    \label{fig:imaging_50kV_1.8mmCu}
\end{figure}

The results of imaging with X-rays of energies $\leq 50$\,keV are shown in Figure~\ref{fig:imaging_50kV_1.8mmCu}. The X-ray tube was placed on the axis of UBAT with voltage (current) set to 50\,kV (5.3\,$\mu$A). The source-detector distance was 8\,090\,mm and a 1.8\,mm thick Cu filter was used to additionally attenuate the flux. For the image reconstruction a conical beam algorithm was used because at this distance the beam illuminating the coded mask does not have exactly parallel rays, but they diverge at an angle of about $\pm0.6^{\circ}$.

Data were processed off-line with analysis software where an FM hit-finder (the same concept as implemented in the UFFO-p electronics) was applied on each 10\,ms data frame in order to reduce the crosstalk and increase the peak signal-to-noise ratio (SNR) in the reconstructed image. The FM hit-finder utilizes a hit pattern recognition technique. The total exposure was 31\,s. The coded mask was attached and it has an open area fraction of about 0.445. There were 4\,026 registered counts due to the background and 4\,781 counts due to the source, and the area of enabled detector pixels was 154.6\,cm$^2$ thus giving a count rate due to the source under the open mask pixels of 2.2 cnt\,s$^{-1}$\,cm$^{-2}$.

\section{Angular Resolution}

\begin{figure}
    \mbox{\includegraphics[width=0.33\linewidth]{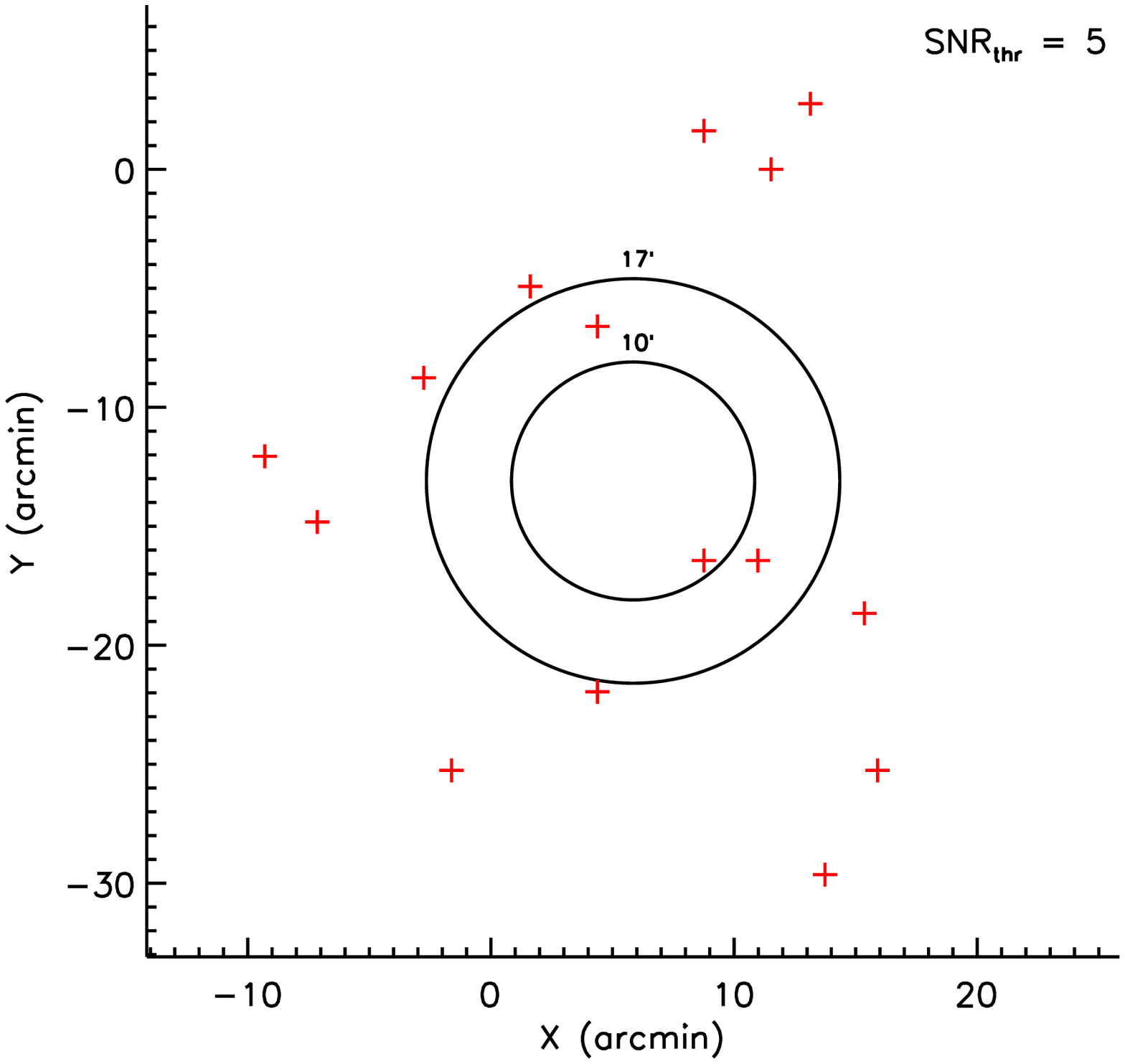}}   
    \mbox{\includegraphics[width=0.33\linewidth]{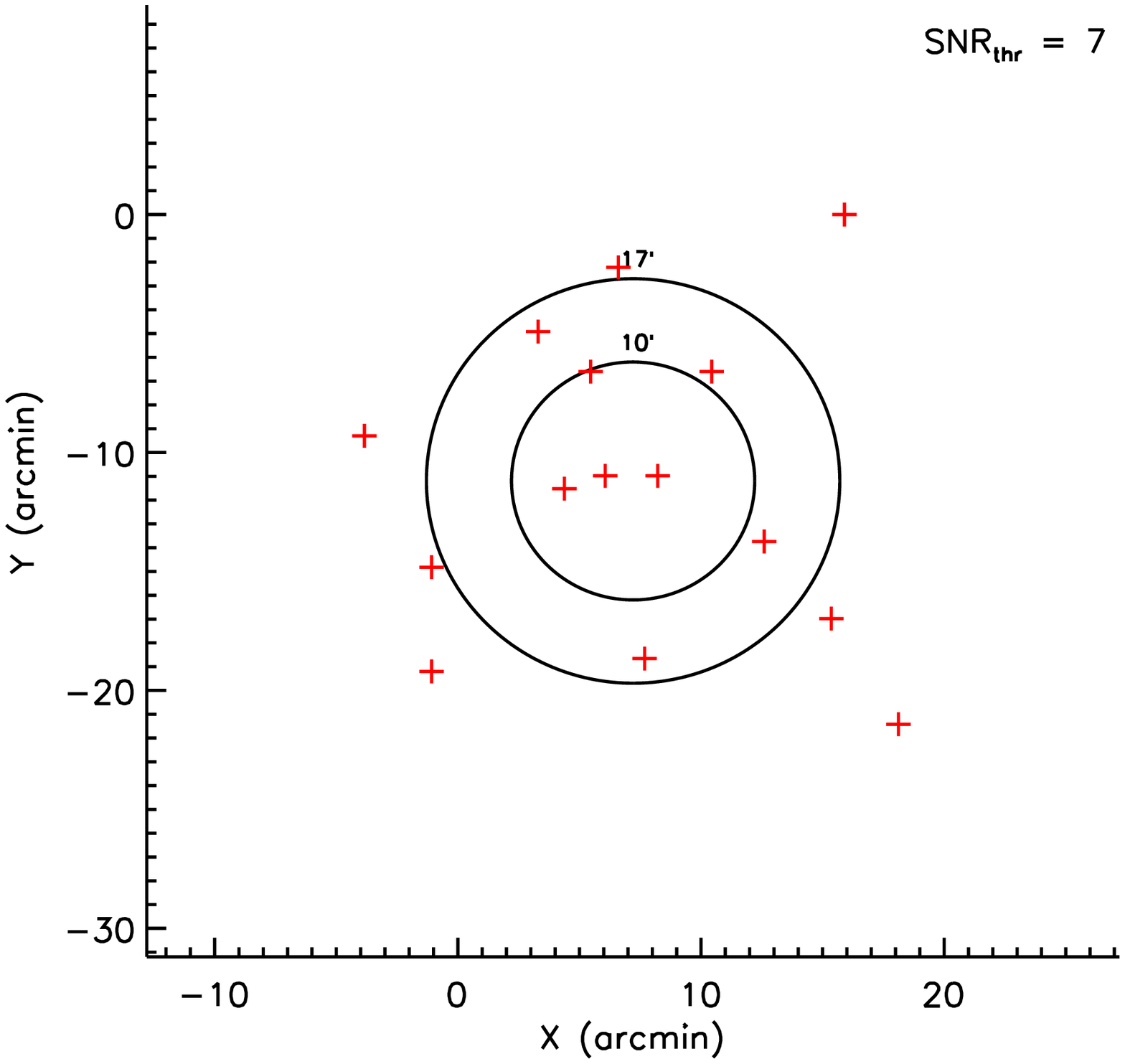}}
    \mbox{\includegraphics[width=0.33\linewidth]{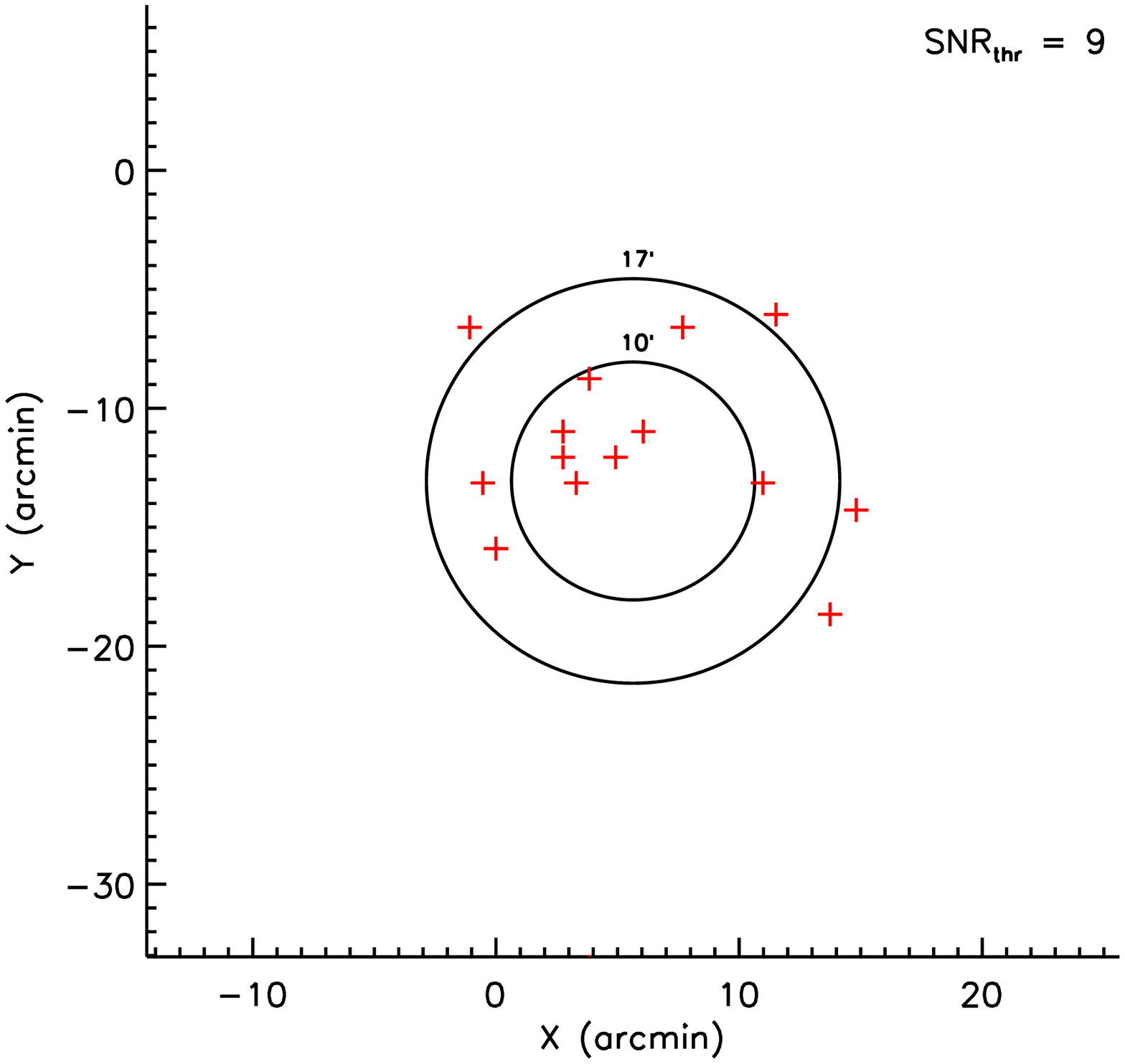}}   
    \caption{The angular resolution of UBAT FM for different detection thresholds is shown. The variance of the measured source positions of 15 different exposures indicate the resolution. SMT has FOV$=17'\times17'$.}
    \label{fig:angular_resolution_method1}
\end{figure}

The X-ray source was placed in a fixed direction, on axis, and the recorded data of one exposure totalling 30\,s were split into 15 independent sub-samples each of duration 2\,s. The imaging algorithm for each sub-sample was run separately. The variance of the calculated directions of the source indicates the angular resolution (Figure~\ref{fig:angular_resolution_method1}). The image reconstruction algorithm was run for three different thresholds of SNR$_{\textrm{thr}}=5, 7, 9$ at which the algorithm was forced to stop. This allows comparison of the dependence of the localization accuracy on SNR. The experimental setup was the same as in the previous section except that here a 1.6\,mm thick Cu filter was used. The count rate due to the source beneath the open mask pixels was 4.1 cnt\,s$^{-1}$\,cm$^{-2}$.

For the detection of SNR\,=\,$5.0\sim5.6$ the standard deviations of the calculated positions in the x-axis and y-axis were $\sigma_{X}=8.2'$ and $\sigma_{Y}=10.3'$.
For SNR\,=\,$7.0\sim7.4$ they were $\sigma_{X}=6.5'$ and $\sigma_{Y}=6.4'$.
For SNR\,=\,$8.6\sim9.4$ they were $\sigma_{X}=5.1'$ and $\sigma_{Y}=6.7'$.
Considering that the angular radius of the source was $1'$, the result is close to the designed resolution of $\pm5\,'$ for $>7\sigma$ detection.

\section{Combined Test with BDRG Detector}
The UBAT FM detector was also tested together with the laboratory version of the Block for X-ray and Gamma-Radiation Detection (BDRG) \cite{Amelyushkin13} scintillator gamma-ray spectrometer. BDRG will be one of the instruments on the Lomonosov satellite. It consists of 75\,mm (diameter) $\times$ 3\,mm (thickness) NaI(Tl) scintillator optically coupled with 80\,mm (diameter) $\times$ 17\,mm (thickness) CsI(Tl). Both crystals are attached to the same photomultiplier tube. The efficiency in the range $30-100$\,keV is close to 100\,\%.

The X-ray tube was set to 50\,kV, 5.3\,$\mu$A, a 1.6\,mm thick Cu filter was placed in front of the tube and the source-detector distance was 7760\,mm. The measured background rate by BDRG in the range $10-100$\,keV was 0.15\,cnt\,s$^{-1}$\,cm$^{-2}$ and the rate due to the source at energy $>$30\,keV was 5.18\,cnt\,s$^{-1}$\,cm$^{-2}$. This count rate can be compared with the measurement done with UBAT FM for the same setup as described in the previous section. However, in that case the source-detector distance was 8090\,mm. The count rate due to the source and under the open mask pixels was 4.10 cnt\,s$^{-1}$\,cm$^{-2}$. Taking into account this distance difference the expected count rate detected by UBAT at distance 7760\,mm would be $\approx 4.46$\,cnt\,s$^{-1}$\,cm$^{-2}$. Comparing this with the count rate from BDRG one can infer the efficiency of UBAT FM at the range of $30-50$\,keV to be $4.46/5.18 = 86\,\%$.


\begin{thebibliography}{99}

\bibitem{Chen11} P. Chen, S. Ahmad, K.~B. Ahn et al., \emph{The UFFO (Ultra Fast Flash Observatory) pathfinder: science and mission}, \emph{Proceedings of the 32nd International Cosmic Ray Conference (ICRC2011), held 11-18 August, 2011 in Beijing, China} {\bf 8} (2011) 243
\bibitem{Nam13} J. Nam, S. Ahmad, K.~B. Ahn et al., \emph{The UFFO Slewing Mirror Telescope for early optical observation from gamma ray bursts}, \emph{Modern Physics Letters A}, {\bf 28} (2013) 40003
\bibitem{Park13} I.~H. Park, S. Brandt, C. Budtz-J{\o}rgensen et al., \emph{Ultra-Fast Flash Observatory for the observation of early photons from gamma-ray bursts}, \emph{New Journal of Physics}, {\bf 15} (2013) 023031
\bibitem{Jung11} A. Jung, S. Ahmad, K.~B. Ahn et al., \emph{Design and fabrication of detector module for UFFO Burst Alert \& Trigger Telescope}, \emph{Proceedings of the 32nd International Cosmic Ray Conference (ICRC2011), held 11-18 August, 2011 in Beijing, China} {\bf 9} (2011), 235
\bibitem{Lee13} J. Lee, S. Jeong, J.~E. Kim et al., \emph{Design, construction and performance of the detector for UFFO Burst Alert \& Trigger Telescope}, in proceedings of \emph{Gamma-ray Bursts: 15 Years of GRB Afterglows}, Edited by A. J. Castro-Tirado, J. Gorosabel, and I. H. Park, \emph{EAS Publication Series}, {\bf 61} (2013), 525
\bibitem{Chang14} Y.-Y. Chang, C.~R. Chen, P. Chen et al., \emph{Inverted-conical light guide for crosstalk reduction in tightly-packed scintillator matrix and MAPMT assembly}, \emph{Nuclear Instruments and Methods in Physics Research Section A}, {\bf 771} (2015), 55
\bibitem{Jeong13} S. Jeong, J.~W. Nam, K.~B. Ahn et al., \emph{Slewing Mirror Telescope optics for the early observation of UV/optical photons from gamma-ray bursts}, \emph{Optics Express}, {\bf 21} (2013), 2263
\bibitem{Kim13} J.~E. Kim, H. Lim, J.~W. Nam et al., \emph{Readout of the UFFO Slewing Mirror Telescope to detect UV/optical photons from Gamma-Ray Bursts}, \emph{Journal of Instrumentation}, {\bf 8} (2013), P07012
\bibitem{Panasyuk11} M. Panasyuk, \emph{Moscow State University satellite "Mikhail Lomonosov" - the multi-purpose observatory in space}, \emph{Proceedings of the 32nd International Cosmic Ray Conference (ICRC2011), held 11-18 August, 2011 in Beijing, China}, {\bf 3} (2011), 313 
\bibitem{Connell13} P.~H. Connell and V. Reglero, \emph{The Ultra Fast Flash Observatory pathfinder - UFFO-p GRB imaging and location with its coded mask X-ray imager UBAT}, in proceedings of \emph{Gamma-ray Bursts: 15 Years of GRB Afterglows}, Edited by A. J. Castro-Tirado, J. Gorosabel, and I. H. Park, \emph{EAS Publication Series}, {\bf 61} (2013), 517
\bibitem{Amelyushkin13} A.~M. Amelyushkin, V.~I. Galkin, B.~V. Goncharov et al., \emph{The BDRG and SHOK instruments for studying gamma-ray burst prompt emission onboard the Lomonosov spacecraft}, \emph{Cosmic Research}, {\bf 51} (2013), 434 

\end{thebibliography}
\end{document}